\documentclass[aps,prd,twocolumn,superscriptaddress]{revtex4-2}

\usepackage{color}

\usepackage{graphicx}
\begin{document}

%\preprint{}

\title{Pre-supernova Ultra-light Axion-like Particles}


\author{Kanji Mori}
\email[]{kanji.mori@fukuoka-u.ac.jp}
%\thanks{}
\affiliation{Research Institute of Stellar Explosive Phenomena, Fukuoka University, 8-19-1 Nanakuma, Jonan-ku, Fukuoka-shi, Fukuoka 814-0180, Japan}
\author{Tomoya Takiwaki}
\affiliation{National Astronomical Observatory of Japan, 2-21-1 Osawa, Mitaka, Tokyo 181-8588, Japan}
\author{Kei Kotake}
%\thanks{}
\affiliation{Research Institute of Stellar Explosive Phenomena, Fukuoka University, 8-19-1 Nanakuma, Jonan-ku, Fukuoka-shi, Fukuoka 814-0180, Japan}
\affiliation{Department of Applied Physics, Faculty of Science, Fukuoka University, 8-19-1 Nanakuma, Jonan-ku, Fukuoka-shi, Fukuoka 814-0180, Japan}


\date{\today}

\begin{abstract}
We calculate the production of ultra-light axion-like particles (ALPs) in a nearby supernova progenitor. Once produced, ALPs escape from the star and a part of them is converted into photons during propagation in the Galactic magnetic field. It is found that the MeV  photon flux that reaches Earth may be detectable by $\gamma$ ray telescopes for ALPs lighter than $\sim1$\,neV when Betelgeuse undergoes oxygen and silicon burning. The dependence of the $\gamma$ ray flux on the stellar mass is much smaller than the uncertainty that originates from the Galactic magnetic field. If ALPs are lighter than $\sim0.1$ neV and the supernova progenitor is close enough to the Solar System, the $\gamma$ ray flux is insensitive to the distance $d$ because the ALP-photon conversion probability is proportional to $d^2$. (Non-)detection of $\gamma$ rays from a supernova progenitor with next-generation $\gamma$ ray telescopes just after pre-supernova neutrino alerts would lead to an independent constraint on ALP parameters as stringent as a SN 1987A limit.
\end{abstract}

%\keywords{}

\maketitle

\section{Introduction}
Axion-like particles (ALPs) are a class of pseudoscalar bosons beyond the standard model of particle physics \citep[e.g.][]{2020PhR...870....1D,2020arXiv201205029C}. They are thought to interact with photons as described by the Lagrangian \citep{1988PhRvD..37.1237R}
\begin{equation}
    \mathcal{L}_{a\gamma}=-\frac{1}{4}g_{a\gamma}F_{\mu\nu}\tilde{F}^{\mu\nu}a,
\end{equation}
where $g_{a\gamma}$ is the coupling constant, $F_{\mu\nu}$ and $\tilde{F}^{\mu\nu}$ are the electromagnetic tensor and its dual, and $a$ is the ALP field. The ALP-photon interaction leads to the production of ALPs in hot astrophysical plasma and ALP-photon conversion in external magnetic fields.

Light ALPs have been pursued experimentally and astronomically because they are a candidate of the dark matter \citep{1983PhLB..120..127P,1983PhLB..120..133A,1983PhLB..120..137D,1992SvJNP..55.1063B,1999NuPhS..72..105K}. Since core-collapse supernovae (SNe) can produce a lot of ALPs via the Primakoff process, a nearby SN can be used as a probe of the ALP mass $m_a$ and the coupling constant $g_{a\gamma}$ \citep{1996PhRvL..77.2372G,1996PhLB..383..439B,2015JCAP...02..006P,2016PhRvD..94h5012F,2017PhRvL.118a1103M}. Since ALPs are converted into photons by the Galactic magnetic field during their propagation, $\gamma$ rays from SNe can be a signature of SN ALPs. Non-detection of $\gamma$ rays from  SN 1987A has provided constraints of $g_{10}=g_{a\gamma}/(10^{-10}\; \mathrm{GeV}^{-1})<0.05$ for $m_a<1$\,neV \citep{2015JCAP...02..006P}. If a future nearby SN is observed by Fermi Large Area Telescope \cite[LAT;][]{2009ApJ...697.1071A}, the SN bound will be significantly improved \citep{2017PhRvL.118a1103M}.

In the previous works on the SN bounds of light ALPs, only ALPs produced after the core-bounce have been discussed. However, in principle, ALPs can be produced in advanced burning phases of massive stars as well because the temperature is sufficiently high. If $\gamma$ ray telescopes are pointed at an SN progenitor following a pre-SN neutrino alert \citep{2004APh....21..303O,2016PhRvD..93l3012Y,2017ApJ...851....6P,2020MNRAS.496.3961K,2020ApJ...899..153M,2020ARNPS..70..121K}, it may be possible to obtain information on ALPs.

The ALP parameters have been explored by ALP helioscopes, haloscopes, and astronomical observations. The CERN Axion Solar Telescope (CAST) obtained a bound $g_{10}<0.66$ for $m_a\lesssim 20$\,meV \citep{2017NatPh..13..584A}. Although the mass range is narrow, the Axion Dark Matter Experiment (ADMX) obtained a very stringent bound of $g_{10}\lesssim10^{-5}$ for $m_a\sim\mu$eV \citep{2020PhRvL.124j1303B}. Searches for $\gamma$ and x ray spectral irregularities of NGC 1275 has provided a constraint of $g_{10}<0.05$ for $m_a<5$\,neV \citep{2016PhRvL.116p1101A} and $g_{10}<(6$--$8)\times10^{-3}$ for $m_a<10^{-3}$\,neV \cite{2020ApJ...890...59R}. Observations of globular clusters have also provided a constraint of $g_{10}<0.66$ for $m_a\lesssim 1$\,keV \citep{2014PhRvL.113s1302A}. Interestingly, ALPs with $g_{10}\gtrsim0.2$ and $m_a\lesssim 0.1\,\mu$eV might be responsible for the observed very-high-energy $\gamma$ ray transparency of the Universe \citep{2013PhRvD..87c5027M}. {Also, a recent observation of the spectra of the cosmic infrared background found excesses which can be interpreted as a hint of photon-ALP mixing \cite{2017PhRvD..96e1701K}.} It has also been pointed out that ultra-light ALPs with $m_a\in[6\times10^{-4},\;10^{-2}]$\,neV are excluded to avoid the superradiance instability of black holes regardless of $g_{a\gamma}$ \citep{2018JCAP...03..043C}.

In this paper, we propose to use nearby SN progenitors to investigate the nature of light ALPs independently of the other methods. There are 31 known SN candidates in the Solar neighborhood within a radius of 1\,kpc, as listed in  Ref.~\citep{2020ApJ...899..153M}. As a benchmark of a nearby SN progenitor, we calculate the $\gamma$ ray flux from $\alpha$ Orionis, which is known as Betelgeuse. The ALP production from Betelgeuse and the resultant x-ray flux have been calculated in previous works \cite{1995PhLB..344..245C,2021PhRvL.126c1101X}. However, the photon spectrum peaks at MeV $\gamma$ ray regions in the last evolutionary stages of massive stars. We discuss the observability of the signature of ALPs by $\gamma$ ray telescopes and possible constraints on ALPs.

This paper is organized as follows. In Section II, we describe our stellar model. In Sections III and IV, we explain the prescription of ALP production rates and the ALP-photon conversion caused by the interaction with the Galactic magnetic field. In Section V, the $\gamma$ ray flux from Betelgeuse observed on Earth is calculated. In Section VI, we provide a fitting formula for the pre-SN ALP spectrum. In Section VII, we study the sensitivity of the $\gamma$ ray flux on the stellar mass and the distance to the star. In Section VIII, the results are summarized and discussed.
\section{Stellar Model}
In order to compute a SN progenitor model, we used Modules for Experiments in Stellar Astrophysics \citep[MESA;][]{Paxton2011,Paxton2013,Paxton2015,Paxton2018,Paxton2019} release 15140. Since the initial mass of Betelguese is estimated to be $18$--$21M_\odot$ \cite{2020ApJ...902...63J}, the initial stellar mass is set to $20M_\odot$. Because of the mass loss \cite{1988A&AS...72..259D,2009A&A...497..255G}, the mass just before the core-collapse is reduced to $16.3M_\odot$. The initial metallicity is set to $Z=Z_\odot=0.0148$ \cite{2019arXiv191200844L}. The calculation is stopped when the infall speed of the iron core reaches 300\,km\,s$^{-1}$.

 \begin{figure}
 \includegraphics[width=8cm]{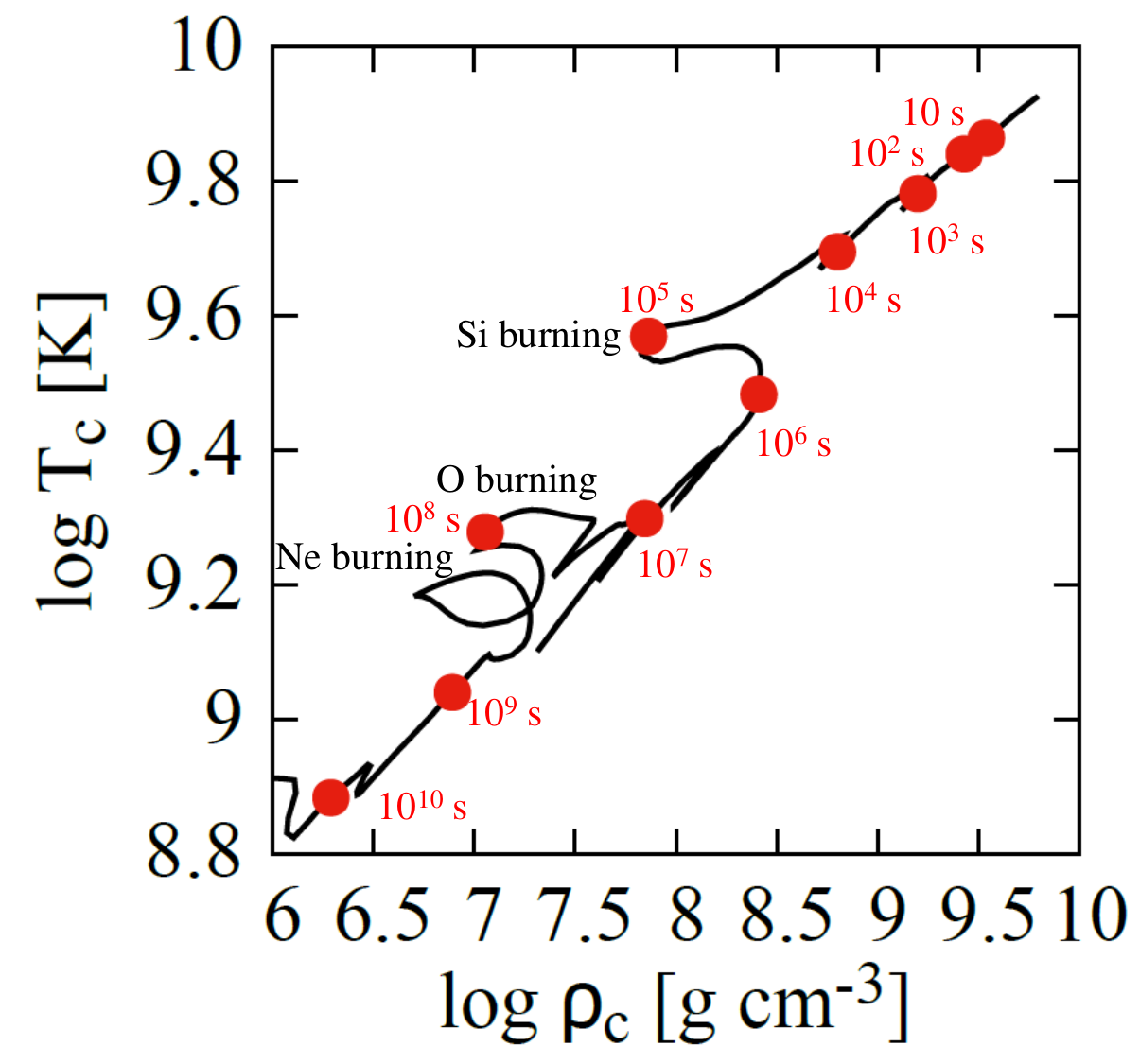}%
 \caption{The evolution of the central temperature $T_\mathrm{c}$ and density $\rho_\mathrm{c}$ in the stellar model. The dots show the time before the core-collapse. \label{evol}}
 \end{figure}
 Fig.~\ref{evol} shows the evolution of the central temperature $T_\mathrm{c}$ and density $\rho_\mathrm{c}$ of the model. The temperature increases as the stellar evolution goes on and exceeds 100\,keV in the neon, oxygen, and silicon burning phases. As shown below, the ALP production in such hot plasma can be so significant that photons that are converted from ALPs by magnetic fields may be observable.
\section{ALP Production Rates}
ALPs are produced by the Primakoff process  induced by the Coulomb field of electrons and ions. The Primakoff rate is given by \citep{2000PhRvD..62l5011D,2020PhLB..80935709C,2020JCAP...12..008L}
\begin{eqnarray}
%\begin{split}
\Gamma_{\gamma\rightarrow a}=g_{a\gamma}^2\frac{T\kappa^2}{32\pi}\frac{p}{E}\left(\frac{((k+p)^2+\kappa^2)((k-p)^2+\kappa^2)}{4kp\kappa^2}\right.\nonumber\\
\left.\times\ln\left(\frac{(k+p)^2+\kappa^2}{(k-p)^2+\kappa^2}\right)
-\frac{(k^2-p^2)^2}{4kp\kappa^2}\ln\left(\frac{(k+p)^2}{(k-p)^2}\right)-1\right),\nonumber\\\label{gamma}
%\end{split}
\end{eqnarray}
where $p$ and $k$ are the momenta of ALPs and photons, respectively, $E$ is the ALP energy, $T$ is the temperature, and
\begin{equation}
\kappa^2=\frac{4\pi\alpha}{T}\frac{\rho}{m_u}\left(R_\mathrm{deg}Y_e+\sum_jZ_j^2Y_j\right)    
\end{equation}
is the Debye-H\"uckel screening scale. Here $\alpha$ is the fine structure constant, $\rho$ is the density, $m_\mathrm{u}$ is the atomic mass unit, $Z_j$ is the atomic number of the $j$-th ion, $Y_e$ and $Y_j$ are the mole fractions of electrons and ions, and $R_\mathrm{deg}$ is the dimensionless degeneracy parameter defined in Ref.~\cite{2019ApJ...881..158S}.
 \begin{figure}
 \includegraphics[width=9cm]{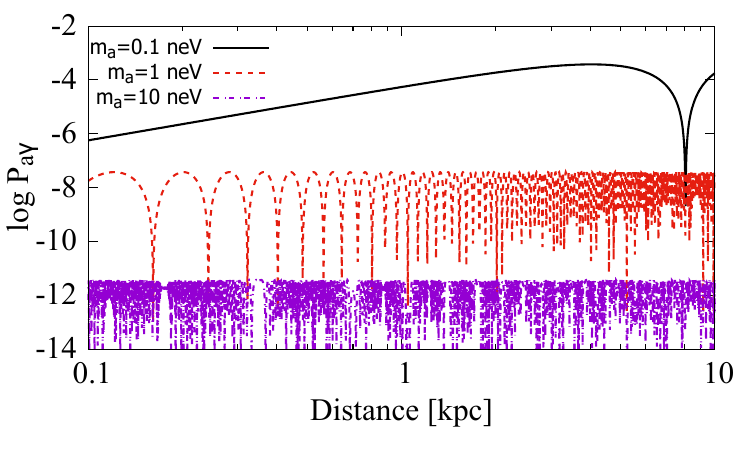}%
 \caption{The ALP-photon conversion probability as a function of the distance $d$ to the SN progenitor. The coupling constant is fixed to $g_{10}=0.05$ and the ALP energy is fixed to $E=1$\,MeV. \label{Pag}}
 \end{figure}

The ALP emissivity per a unit volume is given by
\begin{eqnarray}
    Q_a=2\int\frac{d^3\mathbf{k}}{(2\pi)^3}\Gamma_{\gamma\rightarrow a}\omega f(\omega)=\int_{m_a}^\infty dEE\frac{d^2n_a}{dtdE},
\end{eqnarray}
where $\omega$ is the photon energy, $f(\omega)$ is the Bose-Einstein distribution, and $d^2n_a/dtdE$ is the ALP production rate per a unit volume, time, and energy. The factor 2 in the second term comes from the polarization of photons. The intrinsic ALP luminosity is then calculated as
\begin{equation}
    L_a=4\pi\int_0^RQ_a(r)r^2dr,
\end{equation}
where $R$ is the stellar radius. The ALP luminosity reaches $L_a\sim2\times10^{43}g_{10}^2$ erg s$^{-1}$ just before the core-collapse. This value is almost independent of ALP masses if $m_a<10$\,keV.
\section{ALP-Photon Conversion}
Since they interact with stellar matter only feebly, ALPs escape from the star once produced and propagate in the interstellar space. The interaction between ALPs and the Galactic magnetic field causes the conversion between ALPs and photons \citep{1988PhRvD..37.1237R}. If the magnetic field is constant along the path, the probability for an ALP to be converted into a photon is given by \citep{2015JCAP...02..006P}
\begin{equation}
    P_{a\gamma}=(\Delta_{a\gamma}d)^2\frac{\sin^2\left(\frac{\Delta_\mathrm{osc}d}{2}\right)}{\left(\frac{\Delta_\mathrm{osc}d}{2}\right)^2},
\end{equation}
where $d$ is the distance which the particle has traveled and
\begin{equation}
    \Delta_\mathrm{osc}=\sqrt{(\Delta_a-\Delta_\mathrm{pl})^2+4\Delta_{a\gamma}^2}
\end{equation}
is the oscillation wave number with $\Delta_a=-m_a^2/2E$, $\Delta_\mathrm{pl}=-\omega_\mathrm{pl}^2/2E$,  $\Delta_{a\gamma}=g_{a\gamma}B_\mathrm{T}/2$, $\omega_\mathrm{pl}$ is the plasma frequency,  and $B_\mathrm{T}$ is the transverse component of the magnetic field.

  The distance to Betelgeuse is estimated to be $d=0.168^{+0.027}_{-0.015}$\,kpc \citep{2020ApJ...902...63J}. On the basis of the proximity of the star to the Sun, we assume that the magnetic field along the path of ALPs is constant. The fiducial value of the transverse magnetic field $B_\mathrm{T}$ is fixed to $1.0$ $\mu$G, although the Galactic magnetic field configuration is uncertain. The value of $B_\mathrm{T}$ is dependent on the magnetic field models and can vary between 0.4 and 3.0 $\mu$G \cite{2011ApJ...736...83H,2012ApJ...757...14J,2019MNRAS.486.4275X,2021PhRvL.126c1101X}. Since the amplitude of $P_{a\gamma}$ is approximately proportional to $B_\mathrm{T}^2$, the $\gamma$ ray flux observed on Earth is simply proportional to $B_\mathrm{T}^2$. The uncertainty in the $\gamma$ ray flux that propagates from the model dependence of $B_\mathrm{T}$ is then estimated to be a factor of $0.16$--$9.0$. In order to calculate $\omega_\mathrm{pl}$ in the interstellar gas, the electron number density of $1.1\times10^{-2}$ cm$^{-3}$ \cite{2000ApJ...545..290A} is adopted.
 
 Fig.~\ref{Pag} shows $P_{a\gamma}$ as a function of the distance to the SN progenitor for ALPs with $E=1$ MeV. It is seen that the amplitude of $P_{a\gamma}$ is larger when ALPs are lighter and hence it is easier to search lighter ALPs. It is also notable that the wave length of $P_{a\gamma}$ is shorter for heavier ALPs. This makes the photon flux on Earth sensitive to the distance $d$ between the star and the Solar System.
 \section{Photon Flux on Earth}
 \begin{figure}
 \includegraphics[width=9cm]{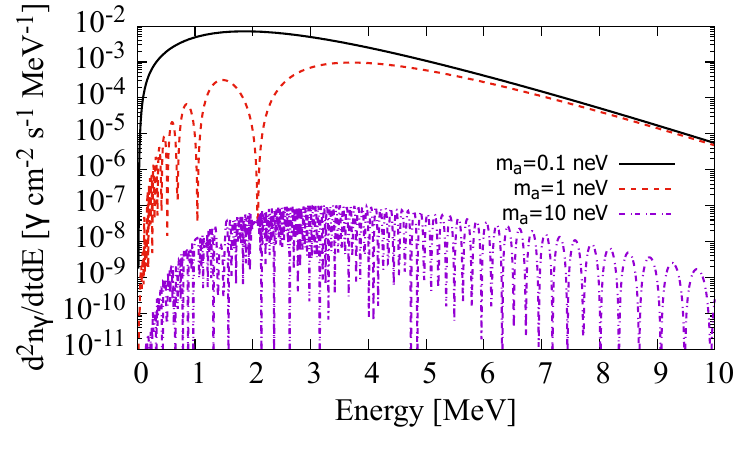}%
 \caption{The spectra of photons which originate from ALPs produced in Betelgeuse just before the core-collapse. The coupling constant is fixed to $g_{10}=0.05$. \label{spec}}
 \end{figure}
 Photons converted from ALPs can be observed by $\gamma$ ray telescopes if its flux is large enough. The photon spectrum observed on Earth is written as
 \begin{equation}
     \frac{d^2n_\gamma}{dtdE}=\frac{1}{4\pi d^2}4\pi P_{a\gamma}\int^R_0 \frac{d^2n_a}{dtdE}r^2dr. \label{gamma_eq}
 \end{equation}
 The spectra of photons that originate from ALPs produced in Betelgeuse just before the core-collapse are shown in Fig.~\ref{spec}. The peak is at 1--5\,MeV and thus the photons converted from ALPs can be a target of MeV $\gamma$ ray telescopes. Since  $P_{a\gamma}$ is dependent on ALP energies, the spectrum is distorted by the factor $P_{a\gamma}$. When ALPs are heavier, the energy dependence of $P_{a\gamma}$ becomes more significant. As a result, the photon spectra for $m_a\gtrsim1$\,neV oscillate as a function of the energy. This spectral irregularity would be a signature of the ALP-photon conversion if it is detected.
 \begin{figure}
 \includegraphics[width=8.5cm]{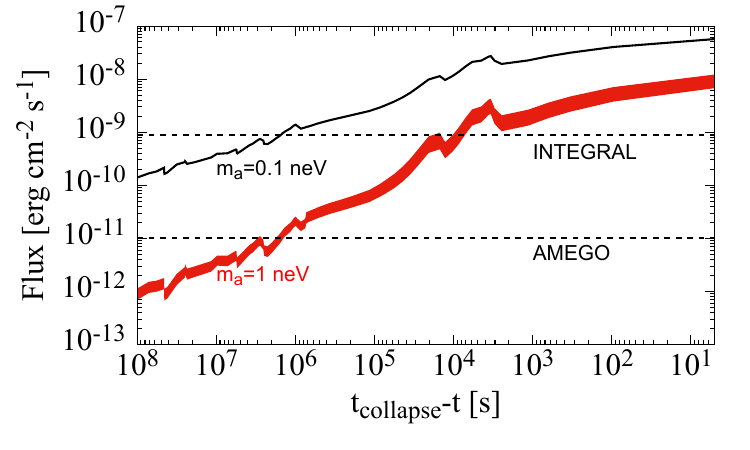}
  \includegraphics[width=8.5cm]{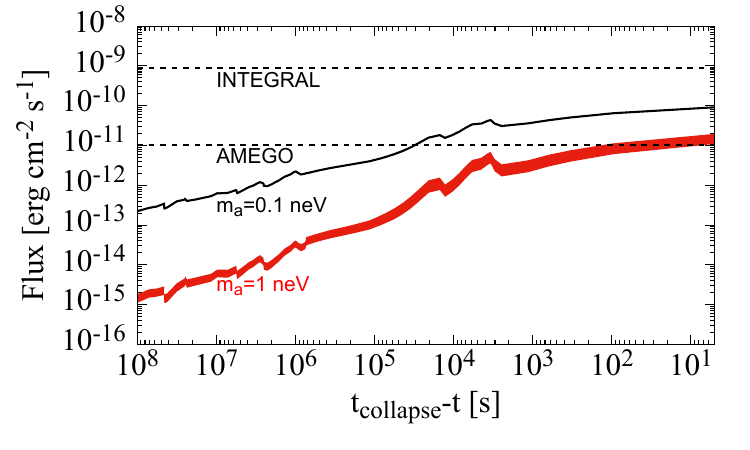}
 \caption{The energy flux of photons which originate from ALPs produced in Betelgeuse as a function of remaining time before the core-collapse. The upper (lower) panel adopts $g_{10}=0.05$ (0.01). The thickness of the curves indicates uncertainties due to the error in the distance to the star. The horizontal lines show the $3\sigma$ sensitivities of INTEGRAL/SPI and AMEGO for 1 MeV photons. \label{Lg}}
 \end{figure}
 
 Fig.~\ref{Lg} shows the energy flux  of photons observed on Earth as a function of time before the core-collapse. The upper and lower panels show the result with $g_{10}=0.05$ and 0.01, respectively. Since $P_{a\gamma}$ is higher for lighter ALPs, the photon flux becomes higher for lower $m_a$. In the case of $g_{10}=0.05$,  the Spectrometer onboard INTEGRAL (SPI) \cite{2003A&A...411L..63V} may observe $\gamma$ rays from Betelgeuse $\sim10^5$ s before its collapse.  Since both of the intrinsic ALP luminosity $L_a$ and the amplitude of $P_{a\gamma}$ are proportional to $g_{a\gamma}^2$, the photon flux is approximately proportional to $g_{a\gamma}^4$. This strong dependence on $g_{a\gamma}$ makes observations of ALPs with low $g_{a\gamma}$ difficult. 
 
 In Fig.~\ref{Lg}, the uncertainty that stems from the distance to Betelgeuse is shown by the thickness of the curves. For ALPs lighter than 1\,neV, the uncertainty is smaller than a factor of 2. However, when ALPs are as heavy as 10\,neV, the wave length of $P_{a\gamma}$ is so short that coherence is lost. As a result, it would be difficult to detect the signature of such ALPs.
 
Observing MeV $\gamma$ rays with high sensitivity is so challenging that instruments more sensitive than the Imaging Compton Telescope \cite[COMPTEL;][]{1993ApJS...86..657S} have not been operated \cite{2004NewAR..48..193S}. Currently, next-generation MeV $\gamma$ ray telescopes including the All-sky Medium Energy Gamma-ray Observatory \cite[AMEGO;][]{2017ICRC...35..798M}, e-ASTROGAM \cite{2018JHEAp..19....1D}, and Sub-MeV/MeV Gamma-ray Imaging Loaded-on-balloon Experiments \cite[SMILE;][]{2021arXiv210700180T} are being developed. These projects aim to achieve higher sensitivity than COMPTEL. The next-generation $\gamma$ ray telescopes are designed to achieve the  sensitivity of $\sim10^{-12}$--$10^{-11}$\,erg cm$^{-2}$ s$^{-1}$ for MeV photons. If these instruments do not detect MeV $\gamma$ rays from Betelgeuse $\sim10^6$ s before its collapse,  the coupling would be constrained to be $g_{10}<0.05$ for $m_a<1$\,neV.
\section{Fitting Formula}
\begin{table}[t]
 \caption{The parameters for the ALP production in pre-SN massive stars. Each parameter is defined in Eq. (\ref{num}).\label{}}
 \begin{ruledtabular}
\begin{tabular}{c|ccc}
$t_\mathrm{collpase}-t$ [s] & $C$              & $E_0$ [MeV] & $\beta$ \\\hline
0                     & $1.68\times10^3$ & 2.54  & 2.50    \\
$10^2$                     & $1.19\times10^3$ & 2.08  & 2.49    \\
$10^3$                     &$9.33\times10^2$            & 1.77  & 2.50    \\
$10^4$                     & $5.98\times10^2$         & 1.57  & 2.47    \\
$10^5$                     & $1.63\times10^2$         & 1.13  & 2.10    \\
$10^6$                     & $2.15\times10^2$         & 0.85  & 2.39    \\
$10^7$                     & $7.31\times10^1$         & 0.61  & 2.10    
\end{tabular}
\end{ruledtabular}
\end{table}
 \begin{figure}
 \includegraphics[width=9cm]{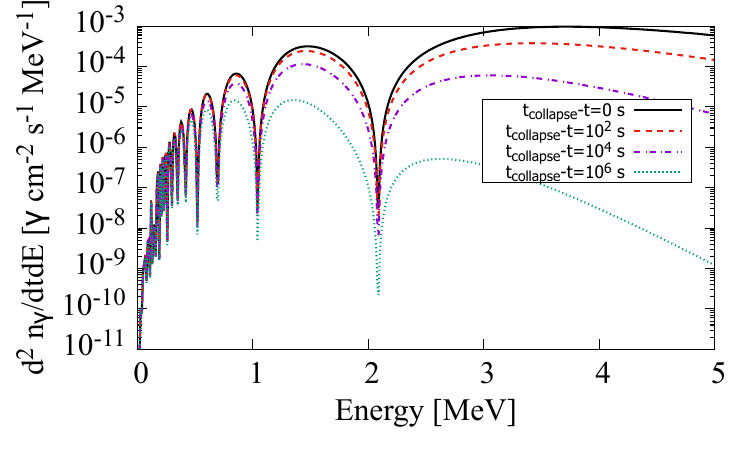}\\%
 \caption{The spectra of photons which originate from ALPs produced in Betelgeuse at 0, $10^2$, $10^4$, and $10^6$\,s before the core-collapse. The ALP mass is fixed to 1\,neV and the coupling is fixed to $g_{10}=0.05$.  \label{spec2}}
 \end{figure}
Recently, Ref.~\cite{2021PhRvL.126c1101X} calculated the ALP production in Betelgeuse and proposed a fitting formula
\begin{equation}
    \frac{d^2N_a}{dtdE}=4\pi\int^R_0\frac{d^2n_a}{dtdE}r^2dr=\frac{10^{47}Cg_{10}^2}{\mathrm{MeV\;s}}\left(\frac{E}{E_0}\right)^\beta e^{-(\beta+1)\frac{E}{E_0}},\label{num}
\end{equation}
where $C$ is the normalization factor, $\beta$ is the spectral index, and $E_0$ is the average energy. The parameters for $t<t_\mathrm{collapse}-1.4$ yr have been tabulated in Ref.~\cite{2021PhRvL.126c1101X} on the basis of their model. In Table I, the parameters which are obtained by fitting the expression to our pre-SN models are shown. It is seen that the average energy shifts from the hard x ray region to the $\gamma$ ray region as time goes on because the temperature becomes higher. Using Eqs. (\ref{gamma_eq}) and (\ref{num}), the photon spectrum on Earth can be evaluated as
\begin{equation}
    \frac{d^2n_\gamma}{dtdE}=\frac{10^{47}Cg_{10}^2P_{a\gamma}}{4\pi d^2}\left(\frac{E}{E_0}\right)^\beta e^{-(\beta+1)\frac{E}{E_0}}\; \mathrm{cm^{-2}\;s^{-1}\;MeV^{-1}}.
\end{equation}
The photon spectra are plotted in Fig.~\ref{spec2}. It is seen that photons become more energetic as a function of time because the temperature becomes higher.
\section{Dependence on Stellar Mass and Distance}
\begin{figure}
  \includegraphics[width=8.5cm]{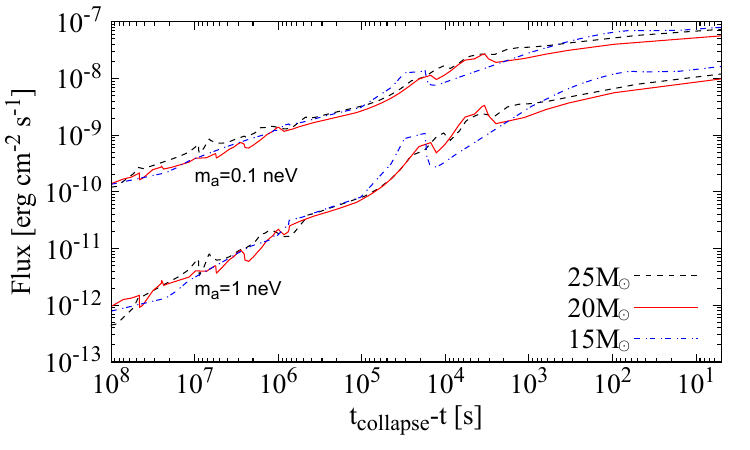}
 \caption{The energy flux of photons which originate from ALPs produced in massive stars with the initial masses of $15M_\odot$, $20M_\odot$ and $25M_\odot$. The ALP masses are fixed to $0.1$ and $1$\,neV. The ALP-photon coupling is fixed to $g_{10}=0.05$. The distance to the stars is assumed to be 0.168\,kpc.  \label{Lgmass}}
 \end{figure}

In the previous sections, we investigated the ALP production in a $20M_\odot$ star. In order to discuss the sensitivity of the stellar mass on the results, we prepared additional stellar models with the initial masses of $15M_\odot$ and $25M_\odot$. The final mass of the $15M_\odot$ model is $13.8M_\odot$ and that of the $25M_\odot$ model is $16.7M_\odot$. Fig.~\ref{Lgmass} shows the $\gamma$ ray flux from the stars. In this figure, the ALP-photon coupling is fixed to $g_{10}=0.05$ and the distance to the stars is assumed to be $0.168$\,kpc, which is equal to the distance to Betelgeuse. It is found that the dependence on the stellar masses is smaller than a factor of three, except for a short period around $\sim10^4$ s before the core-collapse.
 
The $\gamma$ ray flux is also dependent on the distance $d$ to the star. If the ALP-photon conversion probability $P_{a\gamma}$ is constant, the flux would be proportional to $d^{-2}$. In reality, the flux is modified by the oscillations in $P_{a\gamma}$ which are shown in Fig.~\ref{Pag} and deviates from the inverse-square law. We can see from Eq. (6) that $P_{a\gamma}$ is proportional to $d^2$ if $\Delta_\mathrm{osc}d/2\lesssim1$. This implies that the $\gamma$ ray flux becomes independent of $d$ if the star is close enough to the Solar System.

\begin{figure}
  \includegraphics[width=8.5cm]{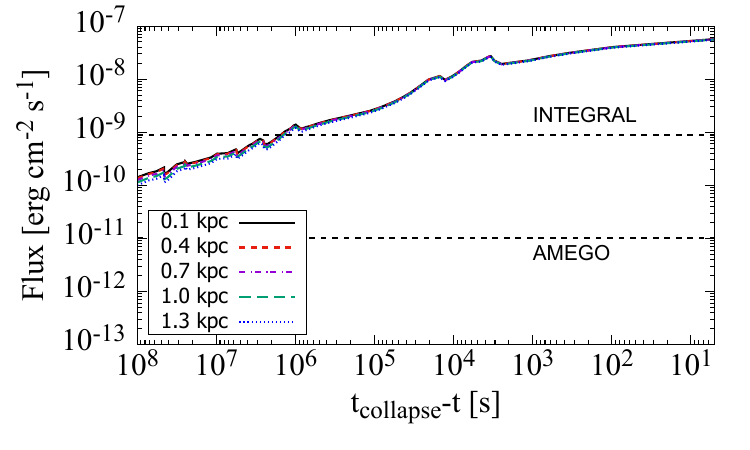}
    \includegraphics[width=8.5cm]{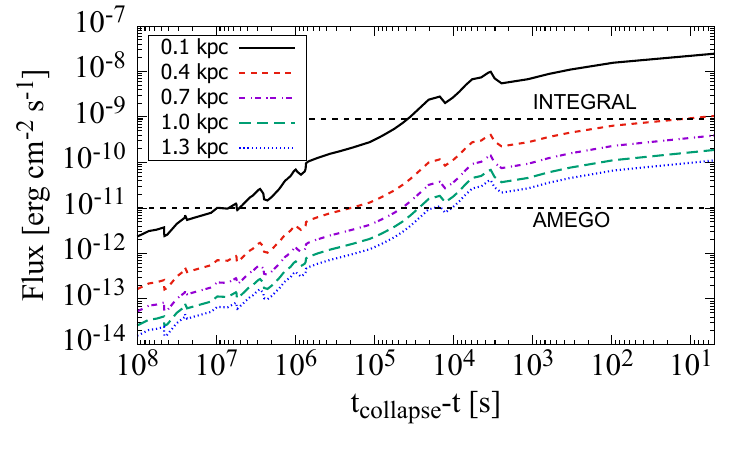}
 \caption{The energy flux of photons which originate from ALPs produced in the $20M_\odot$ model. The ALP masses are fixed to (upper) $0.1$ and (lower) $1$\,neV. The ALP-photon coupling is fixed to $g_{10}=0.05$.  \label{Lgdist}}
 \end{figure}

Fig.~\ref{Lgdist} shows the $\gamma$ ray flux with different distances to the star. The upper panel adopts $m_a=0.1$\,neV and the lower panel adopts $m_a=1$\,neV. In both panels, the $20M_\odot$ model is adopted and the ALP-photon coupling is fixed to $g_{10}=0.05$. In the case of $m_a=0.1$\,neV, the flux decreases as a function of $d$ more slowly than expected from the square-inverse law. This is because $P_{a\gamma}$ is proportional to $d^2$ in this case. As a result, it is possible to obtain a limit as stringent as the SN 1987A limit \citep{2015JCAP...02..006P} with AMEGO even if the star is as far as $\sim1$\,kpc.  On the other hand, in the case of $m_a=1$\,neV, the $\gamma$ ray flux almost follows the inverse-square law because the wave length of $P_{a\gamma}$ is shorter. As a consequence, it would be difficult to obtain a stringent limit on $g_{a\gamma}$ if $d\gtrsim0.5$\,kpc. In the case of $m_a\lesssim0.1$ neV, 31 candidates of pre-SN stars closer than 1\,kpc listed in Ref.~\cite{2020ApJ...899..153M} would be useful to obtain a new constraint on ALPs. Although it may be possible to detect pre-SN $\gamma$ rays from farther stars, the treatment of the ALP-photon conversion across $\gtrsim1$ kpc is beyond the limitation of our model because the magnetic field can deviate from its local value \cite{2011ApJ...736...83H,2012ApJ...757...14J,2019MNRAS.486.4275X}. If the ALP mass is around $m_a \sim 1$\,neV, the number of candidates decreases to 21 if we adopt the threshold of $d=0.5$\,kpc.
 
%%%%%%%%%%%%%%%%%%%%%%%%%%%%%%%%%%%%%%%
 \section{Possible Constraint on ALPs}
\begin{figure}
  \includegraphics[width=8.5cm]{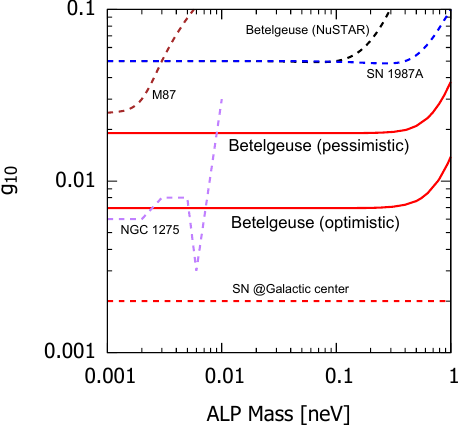}
 \caption{The lower limits on $g_{10}$. The solid line shows the possible limits from pre-SN $\gamma$rays from Betelgeuse. It is assumed that AMEGO observes the star for $10^5$ s before its collapse. The pessimistic limit adopts $B_\mathrm{T}=0.4\;\mu$G and the optimistic limit adopts $B_\mathrm{T}=3.0\;\mu$G. The dotted lines are astrophysical limits that have been obtained so far \cite{2015JCAP...02..006P,2017JCAP...12..036M,2020ApJ...890...59R,2021PhRvL.126c1101X} and a possible limit from an SN at the Galactic center \cite{2017PhRvL.118a1103M}. \label{limit}}
 \end{figure}

%%%%%%%%%%%%%%%%%%%%%%%%%%%%%%%%%%%%%%%
As we have discussed, observations of pre-SN $\gamma$ rays from Betelgeuse would provide information on the ALP parameters. Fig.~\ref{limit} shows possible upper limits on $g_{10}$. It is assumed that AMEGO observes Betelgeuse for $10^5$\,s before its core-collapse since Jiangmen Underground Neutrino Observatory \cite[JUNO;][]{2016JPhG...43c0401A} is expected to announce an early SN alarm a few days before the event \cite{2020ARNPS..70..121K}. We considered the uncertainty of the Galactic magnetic field because it is the most dominant source of uncertainty for the $\gamma$ ray flux. The limit in the pessimistic case assumes $B_\mathrm{T}=0.4\,\mu$G while the optimistic limit  assumes $B_\mathrm{T}=3.0\,\mu$G. In the mass range of $m_a\sim0.01$--$1$\,neV, the pre-SN $\gamma$ ray limit can be the most stringent constraint on $g_{10}$ of the astrophysical limits reported so far, although ALPs from the core-collapse itself would provide an even more stringent limit.
 
%%%%%%%%%%%%%%%%%%%%%%%%%%%%%%%%%%%%%%%
\section{Summary and Discussion}
%%%%%%%%%%%%%%%%%%%%%%%%%%%%%%%%%%%%%%%
In this study, we calculated the ALP production in a pre-SN massive star and the $\gamma$ ray flux that originates from the ALP-photon conversion by the Galactic magnetic field. In particular, we discussed the observability of $\gamma$ rays from Betelgeuse. It was shown that non-detection of $\gamma$ rays from Betelgeuse by next-generation $\gamma$ ray satellites would lead to a constraint $g_{10}<0.05$ for ultra-light ALPs with $m_a<1$\,neV. This constraint is as stringent as the SN 1987A limit \citep{2015JCAP...02..006P}.

We investigated the dependence of the $\gamma$ ray flux on the distance to the SN progenitor and the stellar mass. Since the ALP-photon conversion probability $P_{a\gamma}$ is proportional to $d^2$ when $d\ll2/\Delta_\mathrm{osc}$, the flux is insensitive to $d$ if the star is close enough to the Solar System. In the case of $m_a=0.1$ neV, the flux stays almost constant as a function of $d$ for $d\lesssim4$ kpc. However, if ALPs are heavier, $P_{a\gamma}$ deviates from the $d^2$ law and the flux decreases as a function of $d$. Also, if the initial stellar mass is between $15M_\odot$ and $25M_\odot$, the difference in the flux is much smaller than the uncertainty in the Galactic magnetic field. 

{Since the ALP luminosity from an SN progenitor is lower than that from an SN explosion, the limit that is achievable with the pre-SN signals is weaker than the limit with a nearby SN \cite{2017PhRvL.118a1103M}. However, the temperature and density profiles near a proto-neutron star in an SN is not  known well because of large uncertainties in physics in extreme environments such as nuclear equations of state \cite[e.g.][]{2013ApJ...765...29C,2013ApJ...764...99S,2019PhRvC.100e5802S,2020PhRvL.124i2701Y,2020ApJ...902..150H}, collective neutrino oscillation \cite[e.g.][]{2016NCimR..39....1M,2017PhRvL.118b1101I,2017ApJ...839..132T,2021arXiv210807281N}, and approximations of the neutrino transport \cite[e.g.][]{2019JPhG...46a4001P}. The ignorance on SN microphysics may lead to significant systematic uncertainties on the constraint on ALPs. On the other hand, the structure of massive stars in advanced burning stages is less uncertain. Therefore nearby SN progenitors would provide more robust constraints than SNe.}

Observations of an SN progenitor before its explosion will be enabled by pre-SN alerts from neutrino detectors \cite{2004NJPh....6..114A}. The most sensitive detector for pre-SN neutrinos is JUNO \cite{2016JPhG...43c0401A}, which can detect signals from Betelgeuse a few days before its core-collapse. {Since the inverse $\beta$ decay of protons is insensitive to direction, it is difficult to uniquely identify the SN progenitor with the pre-SN alerts. Nevertheless,  AMEGO would detect signatures of ALPs if $m_a<0.1$\,neV and $g_{10}>0.01$ even without directional information since it plans all-sky surveys with the field of view of 2.5 sr and the cadence of 3 hours. On the other hand, when other $\gamma$ ray telescopes such as INTEGRAL are not performing surveys, it is desirable for them to visit every candidate of the SN progenitor just after the pre-SN alerts. }

{Aside from SN alerts, pre-SN neutrinos would also provide information on the temperature profile in the stellar core. Since stars are optically thin in terms of neutrinos and ALPs, the temperature profile inferred from pre-SN neutrinos would coincide with the environment where ALPs are produced. On the other hand, since the SN core is opaque to neutrinos, we cannot infer the environment at the ALP production site  from SN neutrinos. It is hence expected that nearby SN progenitors provide a unique opportunity to explore ALP production rates.}

The prediction performed in this study is expected to have uncertainties. It has been pointed out that the production of pre-SN neutrinos in massive stars is sensitive to properties of oxygen and silicon shell burning \cite{2016PhRvD..93l3012Y}, of which our understanding is limited. Similarly, the production of ALPs would depend on stellar models. It is hence desirable to perform systematic investigations on model dependence.

Because of difficulties in parallax measurements, the distance to  Betelgeuse is uncertain. However, the uncertainty in the photon flux due to the distance is moderate, as seen in Fig.~\ref{Lg}.  The most important source of uncertainties is the Galactic magnetic field because  $P_{a\gamma}$ is proportional to $B_\mathrm{T}^2$. The uncertainty of the $\gamma$ ray flux due to the magnetic field is estimated to be almost two orders of magnitude. However, because of proximity of the SN progenitors to the Solar System, the uncertainty is smaller than the one which limits from other objects like SN 1987A \citep{2015JCAP...02..006P} and NGC 1275 \citep{2016PhRvL.116p1101A,2020ApJ...890...59R} suffer from. Since each limit suffers from different systematic uncertainties, it is important to explore various methods to explore the ALP parameter space.

\begin{acknowledgments}
This work is supported by Research Institute of Stellar Explosive Phenomena (REISEP) at Fukuoka University and
also by Grants-in-Aid for Scientific Research of the Japan Society for the Promotion of Science (JSPS, Nos. 
JP17H01130, % Kotake Kiban A
JP18H01212, % Yokoi Kiban B
and JP21H01088),% Sotani Kiban B
the Ministry of Education, Science and Culture of Japan (MEXT, No. JP17H06357) % Kotake Shingakujyutu
and JICFuS as “Program for Promoting researches on the Supercomputer Fugaku” (Toward a unified view of 
the universe: from large scale structures to planets, JPMXP1020200109). 
Numerical computations were carried out on PC cluster at Center for Computational Astrophysics, National Astronomical Observatory of Japan.

\end{acknowledgments}

\bibliographystyle{h-physrev}
\bibliography{ref.bib}

\end{document}